\documentclass[12pt]{article}
\usepackage{graphicx}
\begin{document}
Draft for CiSE

\bigskip
\centerline{\bf Simulation of language competition by physicists}

\bigskip

\centerline{Christian Schulze and Dietrich Stauffer}

\section{Introduction}
Physicists have applied their methods to fields outside physics since a long
time: biology, economics, social phenomena, $\dots$. Similar methods sometimes
were also applied by experts from the fields, like Stuart Kauffman for genetics
and Nobel laureates Harry Markowitz and Thomas Schelling for stock markets
and racial segregation, respectively. Thus physicists may re-invent the wheel,
with modifications. However, everybody who followed Formula 1 car races knows
how important slight modifications of the Bridgestone or Michelin tires are.
Biological \cite{brazil}, economic \cite{athana} and social \cite{socio,maxi} 
applications by physicists were already discussed in earlier issue of this 
CiSE department. A common property of many of the simulations by physicists is
that they deal with individuals (atoms, humans, $\dots$) since half a century,
who may be called ``agents'' outside physics.

Thus this review summarises a more recent fashion in physics triggered by
Abrams and Strogatz \cite{abrams}: 
Computer simulation of language competition. The world knows thousands of living
languages, from Chinese spoken by $10^9$ people to many dying languages with 
just one surviving speaker. Can we explain this language size distribution 
$n_s$, Fig.1, by simple models? Will we soon all speak the same language and its
dialects?

First we summarise several models of others, then we present in greater detail
variants of our model \cite{schulze} without asserting that it is always better.

\section{Various models}

\begin{figure}[hbt]
\begin{center}
\includegraphics[angle=-90,scale=0.5]{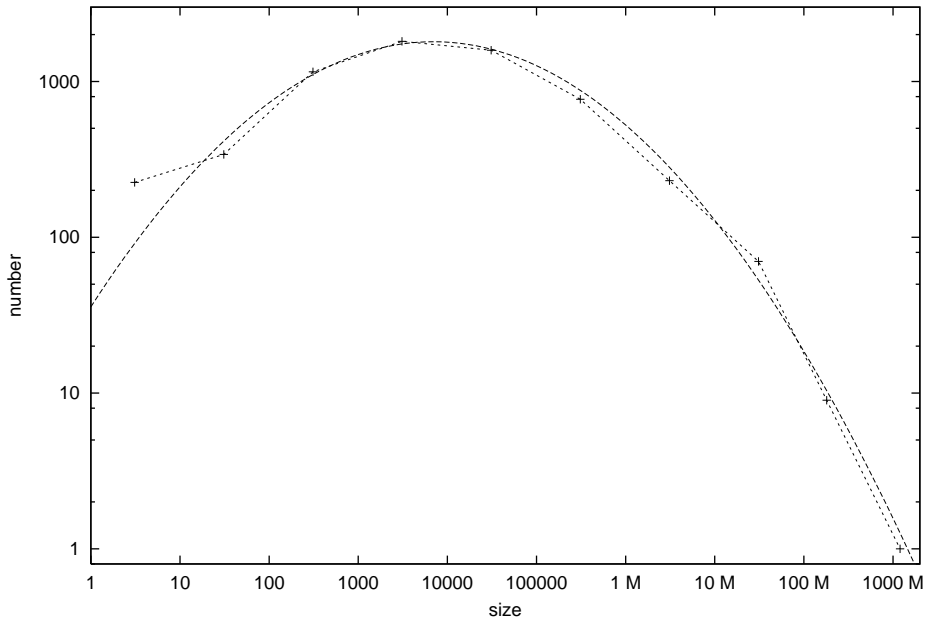}
\end{center}
\caption{Current size distribution $n_s$ of human languages. This size $s$ is 
the number of people having one particular language as mother tongue, and we
plot the number $n_s$ of such languages versus their $s$. Most sizes were binned
to give better statistics. The parabola indicates a log-normal distribution
which fits well except for very small sizes \cite{sutherland,newbook}.}
\end{figure}

\begin{figure}[hbt]
\begin{center}
\includegraphics[angle=-90,scale=0.5]{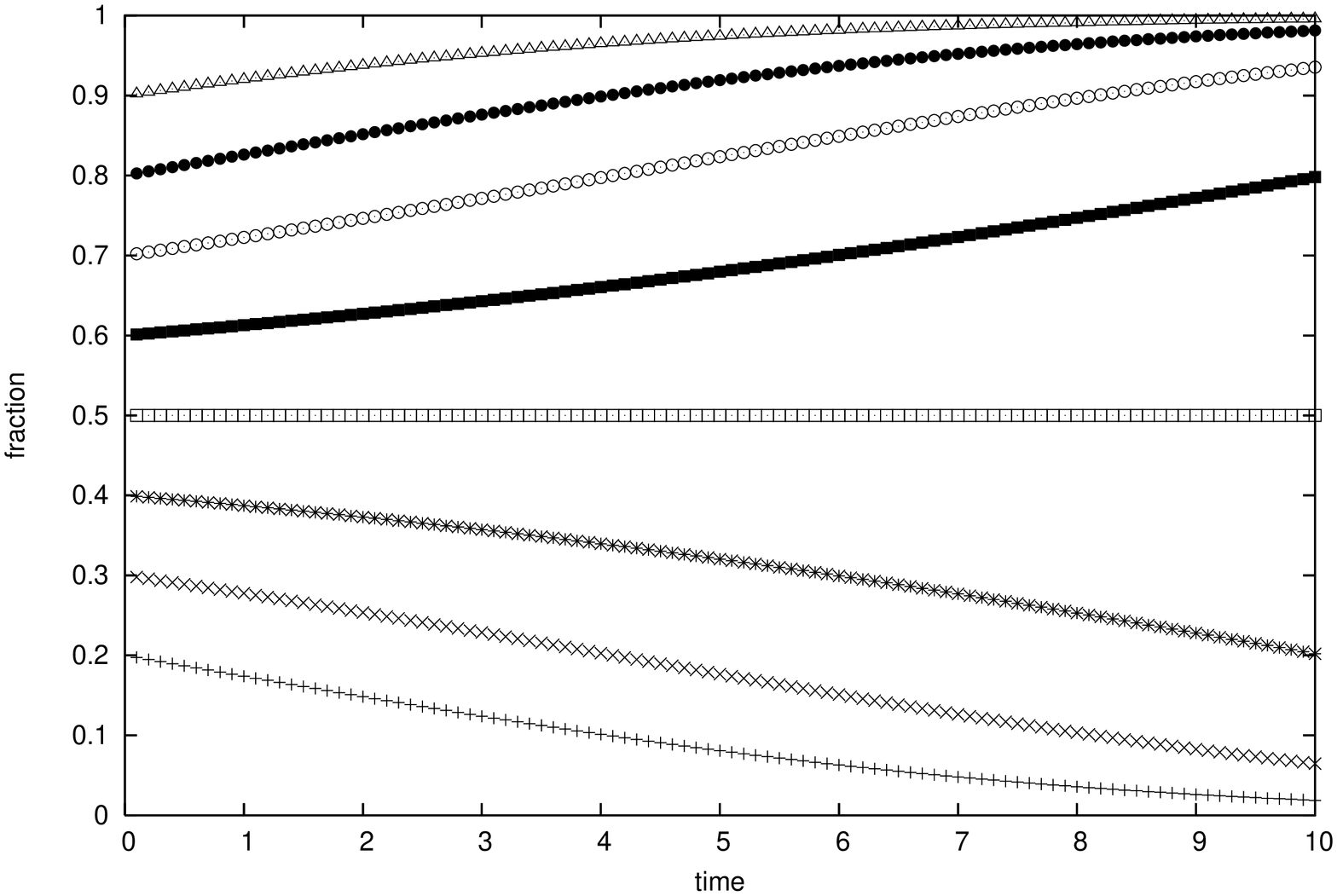}
\end{center}
\caption{Abrams-Strogatz differential equation for two languages: The
initially smaller one faces extinction \cite{abrams,newbook}.}
\end{figure}

The web site http://www.isrl.uiuc.edu/amag/langev/ contains about thousand 
papers on computational linguistics, but mostly on the learning of languages
by children or the evolution of human languages out of simpler sounds a million 
years ago \cite{brighton}. The more recent development seems triggered by 
\cite{abrams} who used differential equations to describe the vanishing of one
language due to the dominance of one other language. Figure 2 shows that
in their basic model the minority language dies out if it does not have a 
higher social status than the majority language. (The populations are in 
principle infinite and thus the fractions in the populations are plotted in 
this Fig.2.) Bilingual people were
included into this model later \cite{spain}, and these differential equations 
on every site of a large lattice allowed to study the coexistence of the
two languages along a rather sharp border \cite{finland}. 

Bit-string models are computationally advantageous, and Kosmidis et al 
\cite{kosmidis} use, for example, 20 bits to describe 20 language features
which can be learned or forgotten by an individual. The first 10 bits 
define one language, the remaining 10 bits another language. In this way,
bilinguals (having nearly 20 bits set) and mixture languages like English
(having ten randomly selected bits set) can be studied as well as monolingual
people (having most of the first 10 or most of the last 10 bits set.) 

Biological ageing was included in \cite{schwammle}; this allowed to take into 
account that foreign languages should be learned when we are young. This work
therefore is a bridge between the language competition reviewed here and 
the older language learning literature of, for example, \cite{nowak}.

All these simulations studied two or a rather limited number of languages. 
A string of $n$ bits instead allows for $2^n$ languages if each different 
bit-string is interpreted as a different language. A two-byte computer word
then corresponds to 16 different important aspects of, say, the grammar of 65536
possible languages \cite{schulze}. 

Also \cite{gomes} simulated numerous languages in a model which allows 
switching from rare to widespread languages, and could reproduce the 
empirical fact that the number of different languages, spoken in an area $A$,
varies roughly as $A^{0.4}$. A linguist and a mathematician 
\cite{wang} had languages defined by strings of integer numbers (similar to
but earlier than our above version) which then allow for branching as in 
biological speciation trees, due to mutations and transfer of numbers.

The above-mentioned model for learning a language \cite{nowak} may also be 
interpreted as a model for competition languages of adults. It uses 
sets of deterministic differential equations and infinite populations, like the 
two-language model of
Abrams and Strogatz \cite{abrams}, but for an arbitrary number of languages.
Starting everybody speaking the same language, one may end up with a 
fragmentation into numerous small languages, due to the natural changes from
one generation to the next. And starting instead with many languages of 
equal size, this situation may become unstable due to these permanent changes
and may lead to the dominance of one language spoken by nearly everybody.
In the style of statistical physics, the many coefficients of the differential 
equations describing the rise and the fall of the many languages can be 
chosen as random instead of making all of them the same \cite{newbook}. Then
Fig.3 shows for the small languages coexisting besides the dominant one a
reasonable size distribution $n_s$. Up to 8000 languages (the current number 
of human languages) were simulated, with two $8000 \times 8000$ matrices for 
the coefficients. Fig.4 shows not the absolute
language sizes but the fractions of people speaking a language. 

In between the learning of a language and the competition of languages is the 
application of the Naming Game \cite{roma}: Two randomly selected  people meet
and try to communicate. The speaker selects an object and retrieves a 
word for it from the own inventory (or invents a new word). The hearer
checks if in the own inventory this object is named by the same word. If 
yes, both players remove from their inventory all other word-object associations
for this object; otherwise the hearer merely adds this association to the own 
inventory. A sharp transition towards understanding was simulated.

\begin{figure}[hbt]
\begin{center}
\includegraphics[angle=-90,scale=0.31]{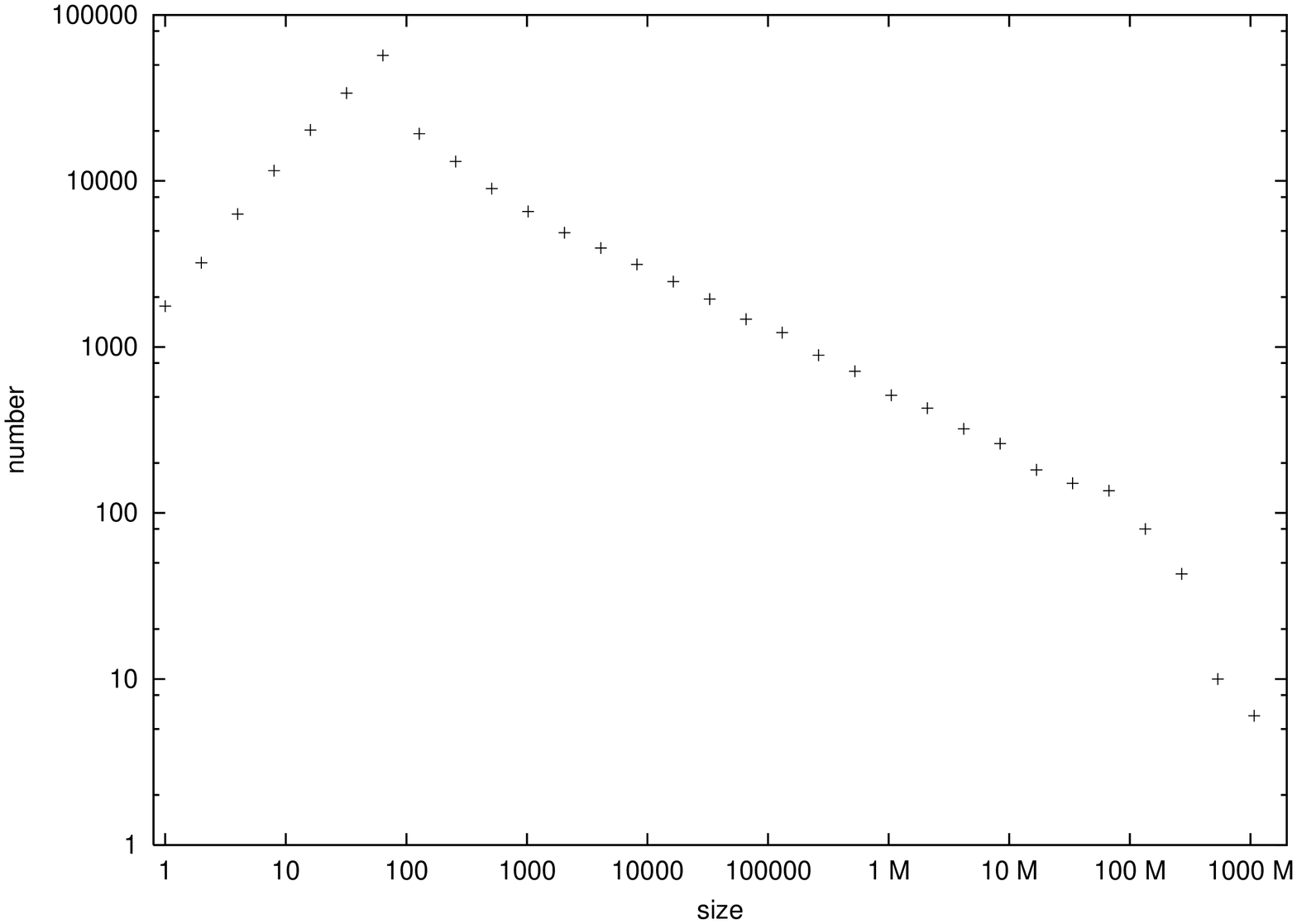}
\includegraphics[angle=-90,scale=0.31]{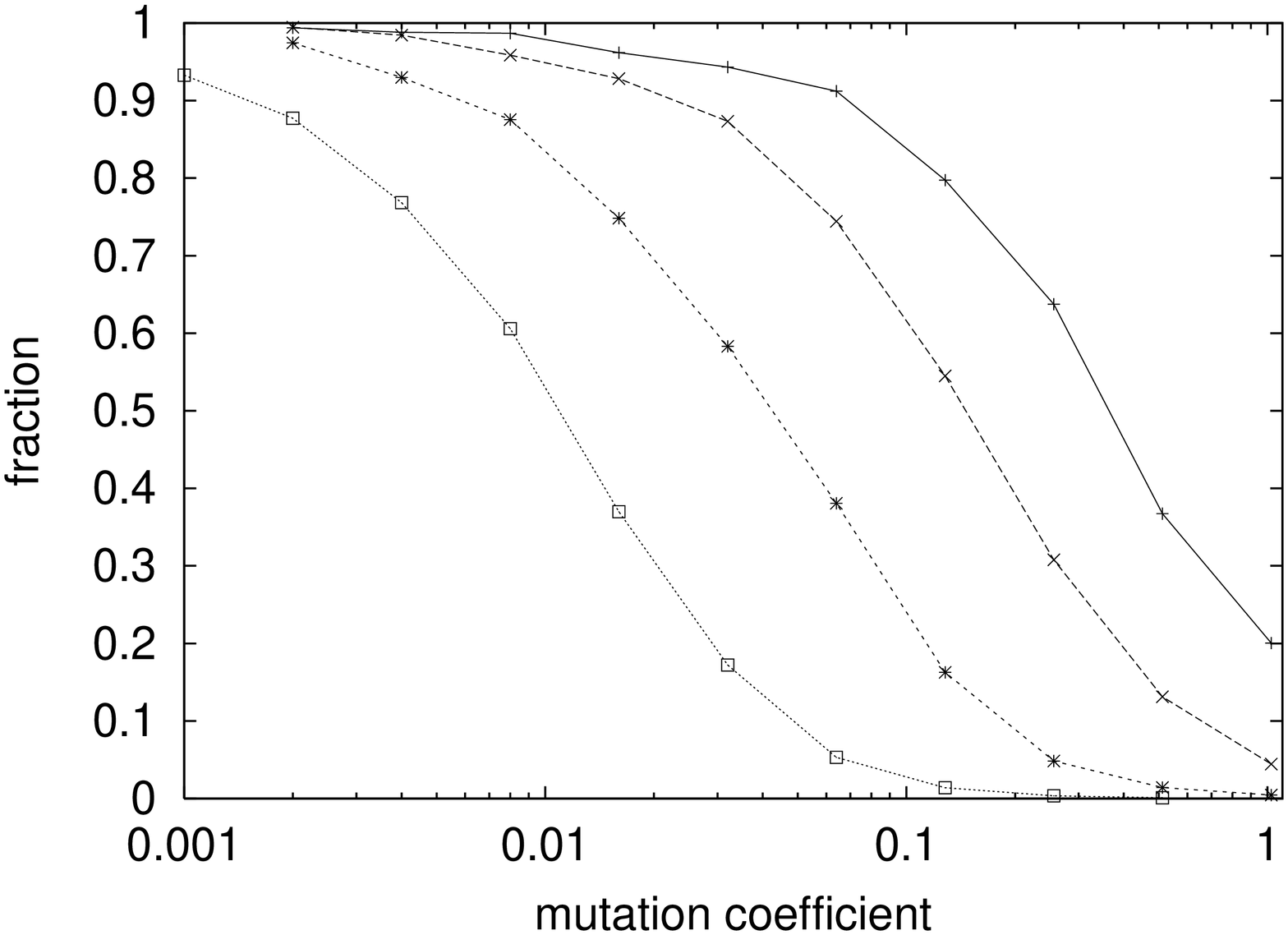}
\end{center}
\caption{Language size distribution $n_s$ (part a) and fraction of people 
speaking the most widespread language (part b), for the model of de Oliveira 
et al \cite{gomes}.}
\end{figure}

If the aim is the reproduce a language size distribution as in Fig.1, then of 
all the models reviewed here the one of \cite{gomes} seems best, since it gave 
language sizes between one and several millions, and thus is explained now  in 
greater detail. The model describes the colonisation of a continent = square 
lattice by people who originally speak one language "zero".  Each
lattice site $k$ has a capability $C_k$ between zero and one, measuring its
resources and proportional to the population on that site. Starting from one
initial site where language $i = 0$ is spoken, in an otherwise empty 
lattice, at each time step one randomly selected empty neighbour of an 
occupied site is occupied, with probability $\propto C_k$. It gets the language
of one of its previously occupied neighbours, with a probability increasing 
with increasing fitness of that neighbour language. This fitness is for each 
language the sum of all $C_k$ for sites speaking this language, i.e. 
proportional to the size of that language.

In order to describe the slow divergence of languages in different regions, such
that not everybody in the whole lattice speaks the same language, the languages
are mutated with probability $\alpha$/fitness, where the mutation coefficient 
$\alpha$ is a free parameter. A mutated language gets an 
integer index $i = 1,2,3, \dots$ not used before for any language. This whole 
process of propagation and mutation stops when all lattice sites are occupied
and the languages are counted. With on average 64 people per site,
and thus $10^{10}$ people per sample, we see in Fig.3a that in many cases
the largest language is spoken by $10^9$ people, as in reality for Chinese.  
Fig.3b shows more systematically the fraction of people speaking the most 
widespread language, as a function of lattice size and mutation coefficient.
We see a smooth transition from domination of one language at small $\alpha$
to fragmentation in many languages at large $\alpha$, with the crossover point
strongly depending on lattice size.
(We summed for Fig.3a over ten lattices of size $16384 \times 16384$ at
a mutation coefficient $\alpha = 0.002$ yielding 20500 languages in each sample.
For Fig.3b we averaged over 100 $L \times L$ lattices with $L = 64$, 256, 1024 
and 4096 from right to left; L = 8192 gave non-monotonic behaviour, not shown.)

\section{Our model}

This section presents a definition of our model \cite{schulze} and then selected
simulations. 

\subsection{Definition}

A language is assumed to be defined by $F$ variables or features,
each of which is an integer between 1 and $Q$. Thus we 
can simulate the competition between up to $Q^F$ different languages. 
In the simplest (bit-string) case we have only binary variables, $Q = 2$.
We assume that on every site of a $L \times L$ square lattice sits exactly one
person, who interacts with the four nearest lattice neighbours. If the 
person dies, his/her child takes the place and speaks the same language,
apart from minor changes which we call ``mutations'' as in biology.

The model is based on three probabilities: the mutation 
probability $p$, the transfer probability $q$, and the flight 
probability $r$. At each of $t$ iterations, each
of the $F$ variables is mutated with probability $p$ to a new
value between 1 and $Q$. This new value is taken randomly with
probability $1-q$, while with probability $q$ it is taken as that of a 
randomly selected lattice neighbour. Finally each person at every iteration,
for each of the $F$ features separately, switches with probability
$(1 - x^2)r$ to the language of another person selected randomly 
from the whole population; here the original language is spoken 
by a fraction $x$ of all people. In this way our model combines 
local and global interactions. Thus languages change continuously through 
$p$ and borrow features from other languages through $q$, while a
small language faces extinction through $r$ because its speakers switch 
to more widespread languages.

This version of our model is both simpler and more complicated than our previous
versions \cite{schulze}. It is more complicated since each feature is an
integer between 1 and $Q$, and not just a bit equal to zero or one. For 
the special case $Q=2$ this does not matter, and then the present version
is simpler since it
has a constant population, thus ignoring the possible human history starting
with one language for Adam and Eve. But the main result will turn out to be the 
same: Starting from one language spoken by everybody, we may see a 
fragmentation into numerous languages, similar to the other biblical story 
of the Babylonian tower. Whether that happens depends on our probabilities for
mutation, transfer and flight $(p,q,r)$.

\begin{figure}[hbt]
\begin{center}
\includegraphics[angle=-90,scale=0.5]{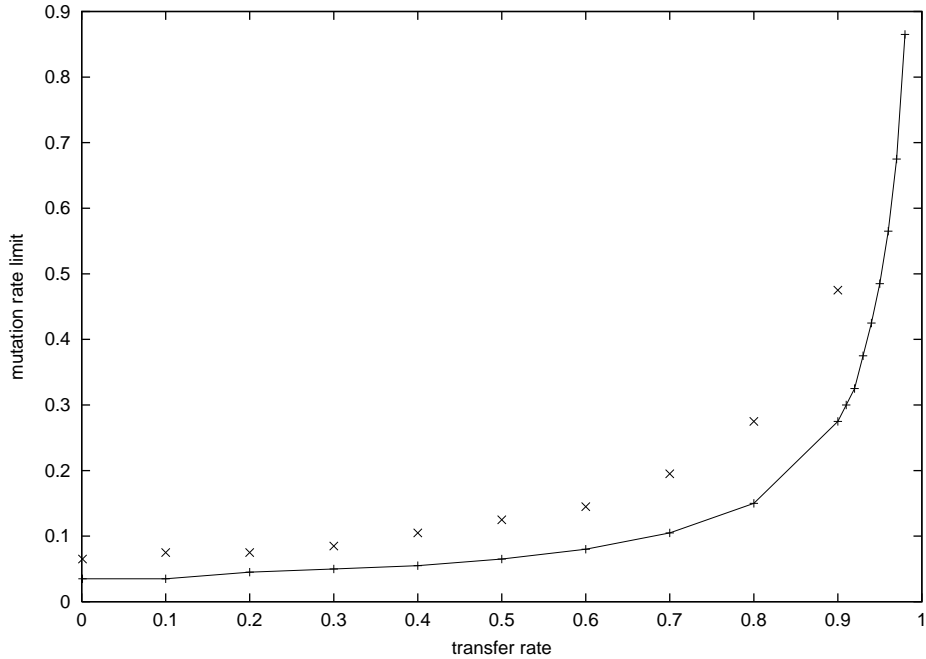}
\end{center}
\caption{Phase diagram: Dominance in the lower right and fragmentation in the 
upper left parts. One sample $L = 301, \; F = 8, \; t = 300.$
The thresholds are about the same for $Q=2,3,5$ and are shown by plus signs 
connected with lines. The single $\times$  signs refer to $Q=10, \; F = 4, \; 
L = 1001$ and agree with those for $Q=3, \; F=4, \; L=301$.}
\end{figure}

\begin{figure}[hbt]
\begin{center}
\includegraphics[angle=-90,scale=0.5]{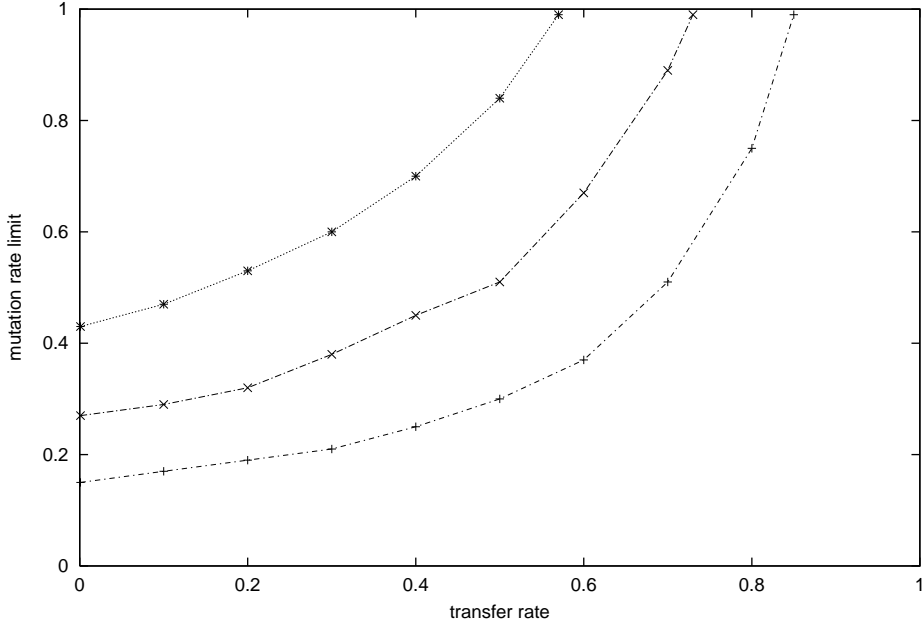}
\end{center}
\caption{As Fig.4 but with single features instead of whole languages switched.
One sample $L = 1001, \; F = 8, \; t = 300; \; Q = 2,3,5$ from bottom to top.}
\end{figure}

\begin{figure}[hbt]
\begin{center}
\includegraphics[angle=-90,scale=0.5]{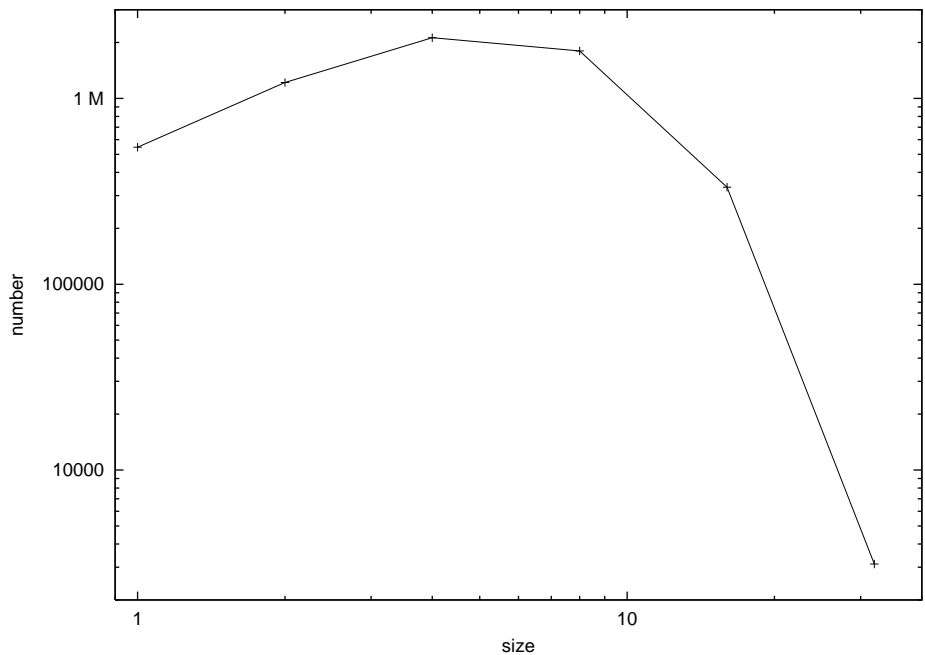}
\end{center}
\caption{Distribution $n_s$ of language sizes, summed from 1000 lattices with $L
= 201$, at $F = 8, \; Q = 3, \; t=300, \; p=q=1/2$.}
\end{figure}

\subsection{Results}

Starting with all variables the same, near $Q/2$, we 
first count how many variables have which value. For small $p$
and large $q$ we found that even after hundreds of iterations
nearly all variables still have their initial values, i.e. the
initial language dominates in the population. For large $p$ and 
small $q$ instead also other values were found in large numbers,
and the population is fragmented into many languages.
Figure 4 shows the phase diagram in the 
$p$-$q$-plane, with dominance everywhere except in the upper 
left corner; for $Q = 2, 3 and 5$ the results roughly agree at $F = 8$, and
also at $F=4$ we get the same curve for $Q=3$ and 10: $F$ is more important 
than $Q$. 
In Fig.5 the switching probability proportional to $r$ is 
defined instead by the size of one feature only, and only this 
feature and not the whole language is switched; now $Q$ is more important.

To get a reasonable size distribution $n_s$ of the shape shown in Fig.1, we 
summed over several lattices and found Fig.6: As in Fig.1 we see a somewhat
asymmetric parabola on this log-log plot, indicating a log-normal distribution
with enhanced small languages. While the shape is good, the absolute sizes
are far too small. In Fig.3a the absolute sizes were nice but the shape was not
rounded enough. Well, you cannot have everything in life.

Our model was continued and improved in \cite{tibihmo} who determined the 
Hamming distance between the languages, which is the number of bits which differ
in a position-by-position comparison of two bit-strings. They then make the 
switching from one language to another depending on the Hamming distance; thus
they took into account that for a Portuguese it is easier to learn Spanish
than Chinese.

\section{Summary} 

We see that lost of recent models have been invented, quite 
independently, at different places in the world. Basically, they can be divided
into those describing two or only few languages, and those who define 
different languages by different bit-strings or generalisations and thus allow 
for numerous possible languages. In the latter case, not necessarily the final
status will be dominance of one languge. Not all these approaches could give 
results like the empirical facts in Fig.1; there is lots of work to be done.

\section{Appendix}

This Fortran 77 program simulates the model of Viviane de Oliveira et al
\cite{gomes} in a memory-saving form. The capacities $C_k$ (= populations
of site $k$) are randomly fixed integers between 1 and 127, and the $limit_k$, 
randomly fixed between 1 and 2047, are the upper limits for the fitness $f_k = 
\sum_k c_k/128$. The $number$ different languages are distinguished
by an index $lang = 1,2,...number$ and are spoken by $icount(lang))$ people.
The language size distribution $ns$ (non-cumulative)
is calculated by binning the sizes in powers of two. The random integers $ibm$
vary between $-2^{63}$ and $+2^{63}$. We follow the Gerling criterion that 
published programs should not contain more lines than authors have years
in their life. Questions should be sent to stauffer@thp.uni-koeln.de.

{\small
\begin{verbatim}
      parameter(L=1024,L2=L*L,L0=1-L,L3=L2+L,L4=25*L+1000,L5=32767)
c     language colonization of de Oliveira, Gomes and Tsang, Physica A 
      dimension neighb(0:3),isite(L0:L3),list(L4),lang(L2),c(L2),f(L5),
     1          ns(0:35),icount(L5),limit(L2)
      integer*8 ibm,mult
      integer*2 lang,limit
      byte isite, c
      data max/2000000000/,iseed/1/,alpha/0.001/,ns/36*0/
      print *, '# ', max, L, iseed, alpha
      mult=13**7
      mult=mult*13**6
      ibm=2*iseed-1
      factor=(0.25d0/2147483648.0d0)/2147483648.0d0
      fac=1.0/128.0
      neighb(0)= 1
      neighb(1)=-1
      neighb(2)= L
      neighb(3)=-L
      do 10 j=2,L5
        icount(j)=0
 10     f(j)=0.0
      do 6 j=L0,L3
        if(j.le.0.or.j.gt.L2) goto 6
        lang(j)=0
 9      ibm=ibm*16807
        c(j)=ishft(ibm,-57)
        if(c(j).eq.0) goto 9
        ibm=ibm*mult
        limit(j)=1+ishft(ibm,-53)
 6      isite(j)=0
      j=L2/2+1
      isite(j)=1
      isite(j+1)=2
      isite(j-1)=2
      isite(j+L)=2
      isite(j+L)=2
      list(1)=j+1
      list(2)=j-1
      list(3)=j+L
      list(4)=j-L
      isurf=4
      nempty=L2-5
      number=1
      lang(j)=1
      icount(1)=1
      f(1)=c(j)*fac
c     surface=2, occupied=1, empty=0
c     end of initialisation, start of growth
      do 1 itime=1,max
        ibm=ibm*16807
        index=1.0+(0.5+factor*ibm)*isurf
        j=list(index)
        if(itime.eq.(itime/50000)*50000)print*,itime,number,isurf,nempty
        ibm=ibm*mult
        if(0.5+factor*ibm .ge. c(j)*fac) goto 1
        list(index)=list(isurf)
        isurf=isurf-1
        isite(j)=1
c       now select language from random neighbour; prob. propto fitness
        fsum=0
        do 5 idir=0,3
 5      if(isite(j+neighb(idir)).eq.1) fsum=fsum+f(lang(j+neighb(idir)))
 3      ibm=ibm*16807
        idir=ishft(ibm,-62)
        i=j+neighb(idir)
        if(isite(i).ne.1) goto 3
        ibm=ibm*mult
        if(0.5+factor*ibm .ge. f(lang(i))/fsum) goto 3
        lang(j)=lang(i)
        f(lang(j))=min(limit(j), f(lang(j)) + c(j)*fac)
c       now come mutations inversely proportional to fitness f
        ibm=ibm*16807
        if(0.5+factor*ibm .lt. alpha/f(lang(j)) ) then
          number=number+1
          if(number.gt.L5) stop 8
          lang(j)=number
          f(lang(j))= c(j)*fac
        end if
        icount(lang(j))=icount(lang(j)) + c(j)
        if(isurf.eq.0) goto 8
c       now determine new surface sites as usual in Eden model
        do 2 idir=0,3
          i=j+neighb(idir)
          if(i.le.0.or.i.gt.L2) goto 2
          if(isite(i).ge.1) goto 2
          isurf=isurf+1
          if(isurf.gt.L4) stop 9
          nempty=nempty-1
          list(isurf)=i
          isite(i)=2
 2      continue
 1    continue
 8    continue
      if(L.eq.79) print 7, lang
 7    format(1x,79i1)
      print *, L, number, itime
      do 11 k=1,number
        j=alog(float(icount(k)))/0.69314
 11     ns(j)=ns(j)+1 
      do 12 j=0,35
 12     if(ns(j).gt.0) print *, 2**j, ns(j)
      stop
      end
\end{verbatim} }

\end{document}